\documentclass[12pt]{article}

\author{Yu.~M.~Zinoviev
       \thanks{E-mail address: ZINOVIEV@MX.IHEP.SU} \\
        {\it Institute for High Energy Physics} \\
        {\it Protvino, Moscow Region, 142284, Russia}}
\title{On Massive Mixed Symmetry Tensor Fields \\
in Minkowski space and (A)dS}

\date{}

\begin{document}

\maketitle

\begin{abstract}
In this paper we give explicit gauge invariant Lagrangian formulation
for massive theories based on mixed symmetry tensors
$\Phi_{[\mu\nu],\alpha}$, $T_{[\mu\nu\alpha],\beta}$ and
$R_{[\mu\nu],[\alpha\beta]}$ both in Minkowski as well as in (Anti) de
Sitter space. In particular, we study all possible massless and partially
massless limits for such theories in (A)dS.
\end{abstract}

\newpage
\setcounter{page}{1}

\section*{Introduction}

In four-dimensional flat Minkowski space-time massive particles are
characterized by one parameter --- spin s and the most simple and
economic description of such particles is the one based on completely
symmetric (spin)-tensors. But moving to the dimensions greater than four,
one faces the fact that representations of appropriate groups
require more parameters and as a result in many interesting cases such
as supergravity theories, superstrings and (supersymmetric) high spin
theories one has to consider different mixed symmetry (spin)-tensors
\cite{Cur86}-\cite{MH02}. In (Anti) de Sitter space the problem becomes even
more complicated because high spin fields in (A)dS reveal a number of
peculiar features such as unitary forbidden regions (i.e. not all
possible values of mass and cosmological constant are allowed) and
appearance of partially massless theories \cite{DN83}-\cite{DW01c}. Moreover,
not all fields admit strictly massless limit \cite{BMV00} making the
very definition of mass for such fields problematic.

In our previous work on this subject \cite{Zin01}  we use gauge invariant
description for massive high spin particles using completely symmetric
tensor fields. Such formulation being unitary and gauge invariant from
the very beginning turns out to be very well suited for the investigation of
unitarity, gauge invariance and partial masslessness. In the present
paper we extended our previous results to the case of mixed symmetry
tensors $\Phi_{[\mu\nu],\alpha}$, $T_{[\mu\nu\alpha],\beta}$ and
$R_{[\mu\nu],[\alpha\beta]}$. In all three cases our strategy will be
as follows. We start with the massless theory in flat Minkowski space
fixing the massless Lagrangian and the structure of gauge transformations.
Then by adding appropriate number of Goldstone fields we construct gauge
invariant formulation for massive particle. Note that gauge transformations
for mixed tensors often turn out to be reducible and the definition of
appropriate set of Goldstone fields is not so trivial as in the case of
symmetric tensors. After that we consider a deformation of the constructed
model to the (A)dS. In contrast with the massless theories our massive
gauge invariant models admit smooth deformation to (A)dS without
introduction any additional fields. At last, having in our disposal massive
theory we study all possible massless as well as partially massless
limits.

\section{$\Phi_{[\mu\nu],\alpha}$ tensor}

Our first example will be the third rank tensor $\Phi_{[\mu\nu],\alpha}$
antisymmetric on the first two indices and satisfying the relation
$\Phi_{[\mu\nu,\alpha]} = 0$ Using these properties it is easy to check
that the following free (quadratic) Lagrangian
\begin{eqnarray}
{\cal L}_0 &=& - \frac{1}{2} \partial^\alpha \Phi^{\mu\nu,\beta}
\partial_\alpha \Phi_{\mu\nu,\beta} + \frac{1}{2} \partial_\alpha
\Phi^{\mu\nu,\alpha} \partial^\beta \Phi_{\mu\nu,\beta} +
\partial_\mu \Phi^{\mu\nu,\alpha} \partial^\beta \Phi_{\beta\nu,\alpha}
+ \nonumber \\
 && + 2 \partial_\alpha \Phi^{\mu\nu,\alpha} \partial_\mu \Phi_\nu +
\partial^\alpha \Phi^\beta \partial_\alpha \Phi_\beta -
(\partial \Phi) (\partial \Phi)
\end{eqnarray}
is invariant under the two gauge transformations
\begin{equation}
\delta \Phi_{\mu\nu,\alpha} = \partial_\mu x_{\nu\alpha} - \partial_\nu
x_{\mu\alpha} + 2 \partial_\alpha y_{\mu\nu} - \partial_\mu y_{\nu\alpha}
+ \partial_\nu y_{\mu\alpha}
\end{equation}
where parameter $x_{\{\alpha\beta\}}$ is symmetric while $y_{[\alpha\beta]}$
--- antisymmetric. Note that these gauge transformations are reducible
in a sense that if one set
\begin{equation}
x_{\alpha\beta} = 3 (\partial_\alpha \xi_\beta + \partial_\beta \xi_\alpha)
\qquad y_{\alpha\beta} = - \partial_\alpha \xi_\beta + \partial_\beta \xi_\alpha
\end{equation}
then $\delta \Phi_{\mu\nu,\alpha} = 0$.

In is not possible to rewrite our Lagrangian as a square of some gauge
invariant quantity because there is no combination of the first
derivatives of $\Phi_{[\mu\nu],\alpha}$ that would be invariant
under both gauge transformations (that will require two derivatives 
\cite{BB02,MH02}). As is rather well known one can however introduce tensor
$T_{[\mu\nu\alpha],\beta}$
\begin{equation}
T_{\mu\nu\alpha,\beta} = \partial_\mu \Phi_{\nu\alpha,\beta} - \partial_\nu
\Phi_{\mu\alpha,\beta} + \partial_\alpha \Phi_{\mu\nu,\beta}
\end{equation}
which is invariant under the $x_{\alpha\beta}$-transformations but not
invariant under the $y_{\alpha\beta}$-ones. Then one rewrite the Lagrangian
in the following simple form:
\begin{equation}
{\cal L}_0 = - \frac{1}{6} T^{\mu\nu\alpha,\beta} T_{\mu\nu\alpha,\beta}
+ \frac{1}{2} T^{\mu\nu} T_{\mu\nu}
\end{equation}
where $T_{[\mu\nu]} = T_{\mu\nu\alpha,}{}^\alpha$.

It is interesting that there exist one more possibility. Namely, one can
introduce another tensor $R_{[\mu\nu],[\alpha\beta]}$
\begin{equation}
R_{\mu\nu,\alpha\beta} = \partial_\alpha \Phi_{\mu\nu,\beta} - \partial_\beta
\Phi_{\mu\nu,\alpha} + \partial_\mu \Phi_{\alpha\beta,\nu} - \partial_\nu
\Phi_{\alpha\beta,\nu}
\end{equation}
which is invariant under the $y_{\alpha\beta}$-transformations but not
under the $x_{\alpha\beta}$-ones and rewrite the same Lagrangian in a
very suggestive form:
\begin{equation}
{\cal L}_0 = - \frac{1}{8} [ R^{\mu\nu,\alpha\beta} R_{\mu\nu,\alpha\beta}
- 4 R^{\mu\nu} R_{\mu\nu} + R^2 ]
\end{equation}

Let us turn now to the massive case. We have two gauge invariances with
the parameters $x_{\alpha\beta}$ and $y_{\alpha\beta}$ so we introduce
two Goldstone fields: symmetric tensor $h_{\{\alpha\beta\}}$ and
antisymmetric one $B_{[\alpha\beta]}$ with their usual kinetic terms:
\begin{eqnarray}
\Delta {\cal L}_0 &=& \frac{1}{2} \partial^\mu h^{\alpha\beta} \partial_\mu
h_{\alpha\beta} - (\partial h)^\mu (\partial h)_\mu + (\partial h)^\mu
\partial_\mu h - \frac{1}{2} \partial^\mu h \partial_\mu h + \nonumber \\
 && + \frac{1}{2} \partial^\mu B^{\alpha\beta} \partial_\mu B_{\alpha\beta}
 + \partial^\mu B^{\alpha\beta} \partial_\alpha B_{\beta\mu}
\end{eqnarray}
and their own gauge transformations:
\begin{equation}
\delta h_{\alpha\beta} = \partial_\alpha x_\beta + \partial_\beta
x_\alpha \qquad
\delta B_{\alpha\beta} = \partial_\alpha y_\beta - \partial_\beta y_\alpha
\end{equation}
With the help of these fields it is easy to check that the sum of massless
Lagrangians supplemented with the following low derivatives terms:
\begin{eqnarray}
{\cal L}_m &=& - m \sqrt{2} (\Phi_{\mu\nu,\alpha} \partial^\mu h^{\nu\alpha}
+ \Phi_\mu (\partial h)^\mu - \Phi_\mu \partial^\mu h ) - \nonumber \\
 && - \frac{m\sqrt{6}}{2} ( \Phi_{\mu\nu,\alpha} \partial^\alpha B^{\mu\nu}
 + 2 \Phi_\mu (\partial B)^\mu ) + \nonumber \\
 && + \frac{m^2}{2} \Phi^{\mu\nu,\alpha} \Phi_{\mu\nu,\alpha} - m^2
 \Phi^\mu \Phi_\mu
\end{eqnarray}
is still invariant under the $x_{\alpha\beta}$, $y_{\alpha\beta}$
transformations provided
\begin{equation}
\delta h_{\alpha\beta} = m \sqrt{2} x_{\alpha\beta} \qquad
\delta B_{\alpha\beta} = m \sqrt{6} y_{\alpha\beta}
\end{equation}

But our two Goldstone fields $h_{\alpha\beta}$ and $B_{\alpha\beta}$
are the gauge fields themselves, so we have to take care about their
own gauge transformations with the parameters $x_\alpha$ and $y_\alpha$.
At first sight it seems that one needs two vector fields to achieve this
goal. But due to reducibility of gauge transformations for
$\Phi_{\mu\nu,\alpha}$ field mentioned above it turns out that it is
enough to introduce only one additional vector fields, the role of the
second one playing the field $\Phi_{\mu\nu,\alpha}$ itself. Indeed, by
introducing vector field $A_\mu$ and adding to the Lagrangian the
following additional terms:
\begin{eqnarray}
\Delta {\cal L} &=& - \frac{1}{4} A_{\mu\nu}{}^2 + m \beta [ h^{\alpha\beta}
\partial_\alpha A_\beta - h (\partial A) + \sqrt{3} B^{\alpha\beta}
\partial_\alpha A_\beta ] - \nonumber \\
 && - m^2 \sqrt{2} \beta \Phi^\mu A_\mu - m^2 \frac{d-1}{d-3} A_\mu{}^2
\end{eqnarray}
we managed not only keep the invariance under the $x_{\alpha\beta}$
and $y_{\alpha\beta}$ transformations, but also to achieve the invariance
under the $x_\alpha$ and $y_\alpha$ transformations, provided:
\begin{eqnarray}
\delta \Phi_{\mu\nu,\alpha} &=& m \alpha [ (g_{\nu\alpha} x_\mu - g_{\mu\alpha}
x_\nu) + \sqrt{3} (g_{\nu\alpha} y_\mu - g_{\mu\alpha} y_\nu)] \nonumber \\
\delta A_\mu &=& m \beta [ x_\mu + \sqrt{3} y_\mu ]
\end{eqnarray}
where $\alpha = - \frac{1}{\sqrt{2(d-3)}}$, $\beta = \sqrt{\frac{d-2}{d-3}}$.
Not that while the structure of massless Lagrangians does not depend on
the dimension of space-time $d$, the structure of massive ones does. In
this section we will assume that $d \ge 4$.

One could note that the vector field $A_\mu$ is also a gauge field and it
seems necessary to introduce one more Goldstone field, namely the scalar
one. Once again it is important to note that the gauge transformations for
$B_{\alpha\beta}$ are also reducible because if one set $y_\alpha =
\partial_\alpha \Lambda$ then $\delta B_{\alpha\beta} = 0$. As a result
one can check that without introduction of any additional fields the
Lagrangian obtained already invariant under the appropriate
transformations which look like:
\begin{equation}
\delta A_\mu = \partial_\mu \Lambda \qquad \delta h_{\alpha\beta} =
\frac{m}{\sqrt{(d-2)(d-3)}} g_{\alpha\beta} \Lambda
\end{equation}

Collecting all pieces together we have the description of massive particle
in terms of four fields $\Phi_{\mu\nu,\alpha}$, $h_{\alpha\beta}$,
$B_{\alpha\beta}$ and $A_\mu$ which is invariant under five gauge
transformations with the parameters $x_{\alpha\beta}$, $y_{\alpha\beta}$,
$x_\alpha$, $y_\alpha$ and $\Lambda$. Note that in $d = 4$ the field
$\Phi_{\mu\nu,\alpha}$ does not describe any physical degrees of freedom,
while the fields $h_{\alpha\beta}$, $A_\mu$ and $B_{\alpha\beta}$ in the
massless limit provide helicities $\pm 2$, $\pm 1$ and 0, respectively.
So in $d = 4$ our theory is just alternative description
of the usual massive spin-2 particle. But in $d > 4$ the field
$\Phi_{\mu\nu,\alpha}$ does introduce additional physical degrees of
freedom, so such theory corresponds to massive representation different
from the one described by usual Fierz-Pauli Lagrangian.

Now let us turn to the (Anti) de Sitter space. We denote covariant
derivative as $D_\mu$ and use the normalization
\begin{equation}
[D_\mu, D_\nu] A_\alpha = R_{\mu\nu,\alpha}{}^\beta A_\beta, \quad
R_{\mu\nu,\alpha\beta} = - \Omega (g_{\mu\alpha} g_{\nu\beta} -
g_{\mu\beta} g_{\nu\alpha}), \quad \Omega =
\frac{2\Lambda}{(d-1)(d-2)}
\end{equation}
where $\Lambda$ --- cosmological constant. As is well known in the (A)dS
even for the massless fields gauge invariance requires that the
non-derivative mass-like terms were present in the Lagrangian. Moreover,
in many cases it is not even possible to have strictly massless limit at
all. So it is convenient to organize the calculations just by the number
of derivatives exactly as in flat space. Then the procedure looks as
follows. We start with the sum of "massless" Lagrangians for all fields
(i.e. the Lagrangians that would describe massless fields in flat space)
\begin{equation}
{\cal L}_0 = {\cal L}_0 (\Phi_{\mu\nu,\alpha}) + {\cal L}_0 (h_{\alpha\beta})
+ {\cal L}_0 (B_{\alpha\beta}) + {\cal L}_0 (A_\mu)
\end{equation}
in which all derivatives are replaced by the covariant ones\footnote{
Note that due to non-commutativity of covariant derivatives there is an
ambiguity because the resulting Lagrangian depends on the order of
derivatives in the initial one. Different choices lead to slightly
different form of mass-like terms, but all choices correspond to
physically equivalent theories.}. Let us consider the gauge
transformations:
\begin{eqnarray}
\delta_0 \Phi_{\mu\nu,\alpha} &=& D_\mu x_{\nu\alpha} - D_\nu x_{\mu\alpha}
+ 2 D_\alpha y_{\mu\nu} - D_\mu y_{\nu\alpha} + D_\nu y_{\mu\alpha}
\nonumber \\
\delta_0 h_{\alpha\beta} &=& D_\alpha x_\beta + D_\beta x_\alpha \qquad
\delta_0 B_{\alpha\beta} = D_\alpha y_\beta - D_\beta y_\alpha \qquad
\delta_0 A_\mu = D_\mu \Lambda
\end{eqnarray}
where all derivatives are also covariant ones. Because the covariant
derivatives do not commute the Lagrangian ${\cal L}_0$ is not invariant 
under such transformations, but as the form of Lagrangian and gauge
transformations is the same as in flat space the residue contains only 
terms with one derivative:
\begin{eqnarray}
\delta_0 {\cal L}_0 &=& - 2 \Omega x^\alpha [ 2 (D h)_\alpha + (d-3) D_\alpha
h ] + \nonumber \\
 && + 2 \Omega x^{\alpha\beta} [ (2d-3) D^\mu \Phi_{\mu\alpha,\beta} + d
 D_\alpha \Phi_\beta - d g_{\alpha\beta} (D \Phi) ] + \\
 && + 3 \Omega y^{\alpha\beta} [ 3 D^\mu \Phi_{\mu\alpha,\beta} - 2
 (d-6) D_\alpha \Phi_\beta ] \nonumber
\end{eqnarray}
These terms do not influent the calculations of variations with two
derivatives, so we keep the same structure of the terms in the Lagrangian
with one derivative as in flat case:
\begin{eqnarray}
{\cal L}_1 &=& - \alpha_1 [ \Phi_{\mu\nu,\alpha} D^\mu h^{\nu\alpha}
+ \Phi_\mu (D h)^\mu - \Phi_\mu D^\mu h ] - \nonumber \\
 && - \frac{\alpha_2}{2} [ \Phi_{\mu\nu,\alpha} D^\alpha B^{\mu\nu} +
 2 \Phi_\mu (D B)^\mu ] + \nonumber \\
 && + \beta_1 [ h^{\alpha\beta} D_\alpha A_\beta - h (D A) ] +
 \beta_2 B^{\alpha\beta} D_\alpha A_\beta
\end{eqnarray}
as well as the form of non-derivative terms in the transformation laws:
\begin{eqnarray}
\delta_1 \Phi_{\mu\nu,\alpha} &=& - \frac{\alpha_1}{2(d-3)} ( g_{\nu\alpha}
x_\mu - g_{\mu\alpha} x_\nu) - \frac{\alpha_2}{2(d-3)} ( g_{\nu\alpha}
y_\mu - g_{\mu\alpha} y_\nu) \nonumber \\
\delta_1 h_{\alpha\beta} &=& \alpha_1 x_{\alpha\beta} + \frac{\beta_1}{d-2}
g_{\alpha\beta} \Lambda \\
\delta_1 B_{\alpha\beta} &=& \alpha_2 y_{\alpha\beta} \qquad \delta_1 A_\mu
= \beta_1 x_\mu + \beta_2 y_\mu \nonumber
\end{eqnarray}
Due to such a choice all the variations with two derivatives cancel each
other and we obtain:
\begin{eqnarray}
\delta_0 {\cal L}_1 + \delta_1 {\cal L}_0 &=& \Omega x^\alpha [ \alpha_1
\frac{d^2-7d +9}{d-3} \Phi_\alpha + 2 \beta_1 (d-1) A_\alpha ] -
\Omega \alpha_2 \frac{d^2-3d+3}{d-3} y^\alpha \Phi_\alpha + \nonumber \\
 && + \Omega \alpha_1 x^{\alpha\beta} (d h_{\alpha\beta} - g_{\alpha\beta} h)
- 3 \Omega \alpha_2 (d-2) y^{\alpha\beta} B_{\alpha\beta}
\end{eqnarray}
Now we add the most general mass-like terms to the Lagrangian:
\begin{equation}
{\cal L}_2 = \frac{c_1}{2} \Phi^{\mu\nu,\alpha} \Phi_{\mu\nu,\alpha}
+ \frac{c_2}{2} \Phi^\mu \Phi_\mu + c_3 \Phi^\mu A_\mu + \frac{c_4}{2}
A_\mu{}^2 + \frac{c_5}{2} h^{\alpha\beta} h_{\alpha\beta} +
\frac{c_6}{2} h^2 + \frac{c_7}{2} B^{\alpha\beta} B_{\alpha\beta}
\end{equation}
and require the cancellation of all variations with one derivative
(taking into account $\delta_0 {\cal L}_0$) and without derivatives
(including $\delta_0 {\cal L}_1 + \delta_1 {\cal L}_0$). This allows one
to express all the parameters in the Lagrangian and the gauge
transformations in terms of $\alpha_1$ and $\alpha_2$:
$$
2c_1 = \alpha_1{}^2 + 2 \Omega(2d-3), \quad
c_2 = - \alpha_1{}^2 - 2 \Omega d, \quad
c_3 = - \alpha_1 \beta_1, \quad c_4 = - \frac{d-1}{d-2} \beta_1{}^2,
$$
$$
c_5 = - \Omega d, \quad c_6 = \Omega, \quad c_7 = 3 \Omega (d-2), \quad
\beta_1 = \sqrt{\frac{d-2}{6(d-3)}} \alpha_2, \quad
\beta_2 = \sqrt{\frac{3(d-2)}{2(d-3)}} \alpha_1
$$
and also gives a very important relation on these two parameters:
\begin{equation}
3 \alpha_1{}^2 - \alpha_2{}^2 + 12 \Omega (d-3) = 0
\end{equation}

Now, having in our disposal massive theory, we can study which
massless or partially massless limits exist in such theory. First of
all, let us note that in the gauge invariant formalism we use the
massless limit means the situation when all Goldstone fields decouple
from the main one. In the case at hand it would requires $\alpha_1 = 0$
and $\alpha_2 = 0$. But the last relation clearly shows that for nonzero
value of cosmological constant it is impossible to have both $\alpha_1 = 0$
and $\alpha_2 = 0$ simultaneously. So, as it was already mentioned in
\cite{BMV00}, there is no fully massless limit for the field
$\Phi_{\mu\nu,\alpha}$ in (A)dS. Instead, depending on the sign of the
cosmological constant, we could obtain one of the two possible partially
massless limits. In AdS ($\Omega < 0$) one can set $\alpha_2 = 0$. As a
result the whole system of four fields decouples into two subsystems.
One of them contains fields $\Phi_{\mu\nu,\alpha}$ and $h_{\alpha\beta}$
with the Lagrangian
\begin{eqnarray}
{\cal L} &=& {\cal L}_0 (\Phi_{\mu\nu,\alpha}) + {\cal L}_0 (h_{\alpha\beta})
- \alpha_1 [ \Phi_{\mu\nu,\alpha} D^\mu h^{\nu\alpha} + \Phi_\mu
(D h)^\mu - \Phi_\mu D^\mu h ] + \nonumber \\
 && + \frac{3\Omega}{2} \Phi^{\mu\nu,\alpha} \Phi_{\mu\nu,\alpha} +
\Omega (d-6) \Phi^\mu \Phi_\mu - \frac{\Omega d}{2} h^{\alpha\beta}
h_{\alpha\beta} + \frac{\Omega}{2} h^2
\end{eqnarray}
where $\alpha_1 = 2 \sqrt{-\Omega(d-3)}$, which is invariant under
the following gauge transformations:
\begin{eqnarray}
\delta \Phi_{\mu\nu,\alpha} &=& D_\mu x_{\nu\alpha} - D_\nu x_{\mu\alpha}
+ 2 D_\alpha y_{\mu\nu} - D_\mu y_{\nu\alpha} + D_\nu y_{\mu\alpha} -
\nonumber \\
 && - \frac{\alpha_1}{2(d-3)} (g_{\nu\alpha} x_\mu - g_{\mu\alpha}
 x_\mu) \\
\delta h_{\alpha\beta} &=& D_\alpha x_\beta + D_\beta x_\alpha +
\alpha_1 x_{\alpha\beta} \nonumber
\end{eqnarray}
As far as we know for the first time such system was considered in
\cite{BMV00}. The rest of the fields ($B_{\alpha\beta}$, $A_\mu$)
gives just the gauge invariant description of massive antisymmetric
tensor with the Lagrangian
\begin{equation}
{\cal L} = {\cal L}_0 (B_{\alpha\beta}) + {\cal L}_0 (A_\mu) +
M B^{\mu\nu} D_\mu A_\nu + \frac{M^2}{4} B^{\mu\nu} B_{\mu\nu}
\end{equation}
which is invariant under:
\begin{equation}
\delta B_{\mu\nu} = D_\mu y_\nu - D_\nu y_\mu \qquad \delta A_\mu =
D_\mu \Lambda + M y_\mu
\end{equation}

On the other hand in the de Sitter space we can set $\alpha_1 = 0$.
Once again the whole system decouples into two subsystems. This time
we obtain partially massless theory with the fields $\Phi_{\mu\nu,\alpha}$
and $B_{\alpha\beta}$ with the Lagrangian
\begin{eqnarray}
{\cal L} &=& {\cal L}_0 (\Phi_{\mu\nu,\alpha}) + {\cal L}_0 (B_{\alpha\beta})
- \frac{\alpha_2}{2} [ \Phi_{\mu\nu,\alpha} D^\alpha B^{\mu\nu} + 2
\Phi_\mu (D B)^\mu ] + \nonumber \\
 && + \frac{\Omega(2d-3)}{2} \Phi^{\mu\nu,\alpha} \Phi_{\mu\nu,\alpha}
 - \Omega d \Phi^\mu \Phi_\mu + \frac{3\Omega(d-2)}{2} B^{\mu\nu} B_{\mu\nu}
\end{eqnarray}
where $\alpha_2 = 2 \sqrt{3\Omega(d-3)}$ and corresponding set of
gauge transformations
\begin{eqnarray}
\delta \Phi_{\mu\nu,\alpha} &=& D_\mu x_{\nu\alpha} - D_\nu x_{\mu\alpha}
+ 2 D_\alpha y_{\mu\nu} - D_\mu y_{\nu\alpha} + D_\nu y_{\mu\alpha} -
\nonumber \\
 && - \sqrt{\frac{3\Omega}{d-3}} (g_{\nu\alpha} y_\mu - g_{\mu\alpha}
 y_\mu) \\
\delta B_{\alpha\beta} &=& D_\alpha y_\beta - D_\beta y_\alpha +
\alpha_2 y_{\alpha\beta} \nonumber
\end{eqnarray}
The rest fields $h_{\alpha\beta}$ and $A_\mu$ with the Lagrangian
\begin{eqnarray}
{\cal L} &=& {\cal L}_0 (h_{\alpha\beta}) + {\cal L}_0 (A_\mu) +
\sqrt{2\Omega(d-2)} [ h^{\alpha\beta} D_\alpha A_\beta - h (D A) ] -
\nonumber \\
 && - \frac{\Omega d}{2} h^{\alpha\beta} h_{\alpha\beta} + \frac{\Omega}{2}
 h^2 - \Omega (d-1) A_\mu{}^2
\end{eqnarray}
and gauge transformations
\begin{equation}
\delta h_{\alpha\beta} = D_\alpha x_\beta + D_\beta x_\alpha +
\sqrt{\frac{2\Omega}{d-2}} g_{\alpha\beta} \Lambda \qquad \delta A_\alpha =
D_\alpha + \sqrt{2\Omega(d-2)} x_\alpha
\end{equation}
is just the gauge invariant description \cite{Zin01} of rather well
known \cite{DN83}-\cite{DW01c} partially massless spin-2 theory in de 
Sitter space.

\section{$T_{[\mu\nu\alpha],\beta}$ tensor}

Our next example --- tensor field $T_{[\mu\nu\alpha],\beta}$
antisymmetric on the first three indices and satisfying the
constraint $T_{[\mu\nu\alpha,\beta]} = 0$. In flat Minkowski space
we will use the following massless Lagrangian
\begin{eqnarray}
{\cal L}_0 &=& \frac{1}{2} \partial^\rho T^{\mu\nu\alpha,\beta} \partial_\rho
T_{\mu\nu\alpha,\beta} - \frac{3}{2} (\partial T)^{\nu\alpha,\beta}
(\partial T)_{\nu\alpha,\beta} - \frac{1}{2} \partial_\beta
T^{\mu\nu\alpha,\beta} \partial^\rho T_{\mu\nu\alpha,\rho} + \nonumber \\
 && + 3 \partial_\beta T^{\mu\nu\alpha,\beta} \partial_\mu T_{\nu\alpha}
 - \frac{3}{2} \partial^\rho T^{\mu\nu} \partial_\rho T_{\mu\nu} +
 3 (\partial T)^\mu (\partial T)_\mu
\end{eqnarray}
where $T_{[\mu\nu]} = T_{\mu\nu\alpha,}{}^\alpha$, which is invariant
under two gauge transformations:
\begin{eqnarray}
\delta T_{\mu\nu\alpha,\beta} &=& 3 \partial_\beta \eta_{\mu\nu\alpha}
+ \partial_\alpha \eta_{\mu\nu\beta} + \partial_\mu \eta_{\nu\alpha\beta}
- \partial_\nu \eta_{\mu\alpha\beta} + \nonumber \\
 && + \partial_\mu \chi_{\nu\alpha,\beta} - \partial_\nu
\chi_{\mu\alpha,\beta} + \partial_\alpha \chi_{\mu\nu,\beta}
\end{eqnarray}
where parameter $\eta_{\mu\nu\alpha}$ completely antisymmetric on all
indices, while $\chi_{\mu\nu,\alpha}$ --- mixed tensor
antisymmetric on first two indices and satisfying $\chi_{[\mu\nu,\alpha]}
= 0$. For what follows it is important to note that these gauge
transformations are also reducible, namely if one set
\begin{eqnarray}
\chi_{\mu\nu,\alpha} &=& \partial_\mu x_{\nu\alpha} - \partial_\nu
x_{\mu\alpha} + 2 \partial_\alpha y_{\mu\nu} - \partial_\mu y_{\nu\alpha}
+ \partial_\nu y_{\mu\alpha} \nonumber \\
\eta_{\mu\nu\alpha} &=& - \frac{1}{2} (\partial_\mu y_{\nu\alpha} - \partial_\nu
y_{\mu\alpha} + \partial_\alpha y_{\mu\nu})
\end{eqnarray}
where $x_{\alpha\beta}$ is symmetric and $y_{\alpha\beta}$ is antisymmetric,
then $T_{\mu\nu\alpha,\beta}$ remains invariant.

To obtain gauge invariant description of corresponding massive field we
introduce two Goldstone fields $\Phi_{\mu\nu,\alpha}$ and $C_{\mu\nu\alpha}$
with the same symmetry properties as $\chi_{\mu\nu,\alpha}$ and
$\eta_{\mu\nu\alpha}$, respectively. The kinetic terms for these fields:
\begin{equation}
\Delta {\cal L} = {\cal L}_0 (\Phi_{\mu\nu,\alpha}) - \frac{1}{2}
\partial^\beta C^{\mu\nu\alpha} \partial_\beta C_{\mu\nu\alpha} +
\frac{3}{2} (\partial C)^{\mu\nu} (\partial C)_{\mu\nu}
\end{equation}
where ${\cal L}_0 (\Phi_{\mu\nu,\alpha})$ is the same Lagrangian as we
use in the previous section. Both fields have its own gauge symmetries:
\begin{eqnarray}
\delta \Phi_{\mu\nu,\alpha} &=& \partial_\mu x_{\nu\alpha} - \partial_\nu
x_{\mu\alpha} + 2 \partial_\alpha y_{\mu\nu} - \partial_\mu y_{\nu\alpha}
+ \partial_\nu y_{\mu\alpha} \nonumber \\
\delta C_{\mu\nu\alpha} &=& \partial_\mu z_{\nu\alpha} - \partial_\nu
z_{\mu\alpha} + \partial_\alpha z_{\mu\nu}
\end{eqnarray}
with $x_{\alpha\beta}$ symmetric, while $y_{\alpha\beta}$ and $z_{\alpha\beta}$
antisymmetric on their indices.

By straightforward calculations one can easily check that with the addition
of the following low derivatives terms
\begin{eqnarray}
{\cal L}_1 &=& m \sqrt{3} [ T_{\mu\nu\alpha,\beta} \partial^\mu
\Phi^{\nu\alpha,\beta} - T_{\mu\nu} \partial_\alpha \Phi^{\mu\nu,\alpha}
- 2 T_{\mu\nu} \partial^\mu \Phi^\nu ] + \nonumber \\
 && + \frac{2m}{\sqrt{3}} [ T_{\mu\nu\alpha,\beta} \partial^\beta
 C^{\mu\nu\alpha} - 3 T_{\mu\nu} (\partial C)^{\mu\nu} ] - \nonumber \\
 && - \frac{m^2}{2} [ T^{\mu\nu\alpha,\beta} T_{\mu\nu\alpha,\beta}
- 3 T^{\mu\nu} T_{\mu\nu} ]
\end{eqnarray}
the whole Lagrangian remains to be invariant under the $\chi_{\mu\nu,\alpha}$
and $\eta_{\mu\nu\alpha}$ transformations provided the Goldstone
fields are transformed as foolows:
\begin{equation}
\delta \Phi_{\mu\nu,\alpha} = m \sqrt{3} \chi_{\mu\nu,\alpha} \qquad
\delta C_{\mu\nu\alpha} = 2 m \sqrt{3} \eta_{\mu\nu\alpha}
\end{equation}

But our Goldstone fields are the gauge fields themselves, so one has to
take care about their own gauge symmetries with the parameters
$x_{\alpha\beta}$, $y_{\alpha\beta}$ and $z_{\alpha\beta}$. At first
sight it seems that we need three more Goldstone fields one symmetric
second rank tensor and two antisymmetric ones. But due to reducibility
of gauge transformations for the field $T_{\mu\nu\alpha,\beta}$ it turns
out to be enough to introduce only one additional field, namely
antisymmetric tensor $B_{[\mu\nu]}$. Indeed with the addition to the
Lagrangian the following new terms
\begin{eqnarray}
\Delta {\cal L} &=& \frac{1}{2} \partial^\mu B^{\alpha\beta} \partial_\mu
B_{\alpha\beta} + \partial^\mu B^{\alpha\beta} \partial_\alpha B_{\beta\mu}
- \nonumber \\
 && - m \sqrt{\frac{d-3}{d-4}} [ \Phi_{\mu\nu,\alpha} \partial^\alpha
 B^{\mu\nu} + 2 \Phi_\mu (\partial B)^\mu + 2 C_{\mu\nu\alpha}
 \partial^\mu B^{\nu\alpha} ] + \nonumber \\
 && + m^2 [ - \sqrt{\frac{3(d-3)}{d-4}} T^{\mu\nu} B_{\mu\nu} +
 \frac{d-2}{d-4} B^{\mu\nu} B_{\mu\nu} ]
\end{eqnarray}
we managed not only to keep invariance under the $\chi_{\mu\nu,\alpha}$
and $\eta_{\mu\nu\alpha}$ transformations, but also achieve the invariance
under the transformations $x_{\alpha\beta}$, $y_{\alpha\beta}$ and
$z_{\alpha\beta}$ as well, provided
\begin{eqnarray}
\delta T_{\mu\nu\alpha,\beta} &=& \frac{2m}{\sqrt{3}(d-4)} [ g_{\alpha\beta}
y_{\mu\nu} - g_{\nu\beta} y_{\mu\alpha} + g_{\mu\beta} y_{\nu\alpha}
+ \nonumber \\
 && + g_{\alpha\beta} z_{\mu\nu} - g_{\nu\beta} z_{\mu\alpha} +
g_{\mu\beta} z_{\nu\alpha} \\
\delta B_{\alpha\beta} &=& 2 m \sqrt{\frac{d-3}{d-4}} (y_{\alpha\beta} -
z_{\alpha\beta}) \nonumber
\end{eqnarray}
One can see that our construction works for $d > 4$ only, because
in $d=4$ the trace part of the $T_{\mu\nu\alpha,\beta}$ completely
decouples in the massless Lagrangian. So we will assume that $d \ge 5$.

It is still not the end of the story because our new Goldstone field
$B_{\alpha\beta}$ is a gauge field itself. As we have already mentioned
in the previous section gauge transformations for the field
$\Phi_{\mu\nu,\alpha}$ are also reducible, as a result there is no need
to introduce any new fields. Indeed, it easy to check that the Lagrangian
obtained so far already invariant under one more gauge transformation
with vector parameter $z_\mu$ having the form:
\begin{equation}
\delta B_{\alpha\beta} = \partial_\alpha z_\beta - \partial_\beta z_\alpha
\qquad \delta \Phi_{\mu\nu,\alpha} = - \frac{m}{\sqrt{(d-3)(d-4)}}
(g_{\nu\alpha} z_\mu - g_{\mu\alpha} z_\nu)
\end{equation}

Thus we have full massive theory with four fields $T_{\mu\nu\alpha,\beta}$,
$\Phi_{\mu\nu,\alpha}$, $C_{\mu\nu\alpha}$ and $B_{\mu\nu}$ which is
invariant under the six gauge transformations with parameters
$\chi_{\mu,\alpha\beta}$, $\eta_{\mu\nu\alpha}$, $x_{\alpha\beta}$,
$y_{\alpha\beta}$, $z_{\alpha\beta}$ and $z_\alpha$. Let us turn now
to (A)dS. We will follow the same procedure as in the previous case
and start with the sum of (covariantized) "massless" Lagrangians
for all four fields
$$
{\cal L}_0 = {\cal L}_0 (T_{\mu\nu\alpha,\beta}) +
{\cal L}_0 (\Phi_{\mu\nu,\alpha}) + {\cal L}_0 (C_{\mu\nu\alpha}) +
{\cal L}_0 (B_{\mu\nu})
$$
as well as the following initial gauge transformations:
\begin{eqnarray}
\delta_0 T_{\mu\nu\alpha,\beta} &=& 3 D_\beta \eta_{\mu\nu\alpha}
+ D_\alpha \eta_{\mu\nu\beta} + D_\mu \eta_{\nu\alpha\beta}
- D_\nu \eta_{\mu\alpha\beta} + \nonumber \\
 && + D_\mu \chi_{\nu\alpha,\beta} - D_\nu
\chi_{\mu\alpha,\beta} + D_\alpha \chi_{\mu\nu,\beta} \nonumber \\
\delta_0 \Phi_{\mu\nu,\alpha} &=& D_\mu x_{\nu\alpha} - D_\nu
x_{\mu\alpha} + 2 D_\alpha y_{\mu\nu} - D_\mu y_{\nu\alpha}
+ D_\nu y_{\mu\alpha} \\
\delta_0 C_{\mu\nu\alpha} &=& D_\mu z_{\nu\alpha} - D_\nu
z_{\mu\alpha} + D_\alpha z_{\mu\nu} \nonumber \\
\delta_0 B_{\alpha\beta} &=& D_\alpha z_\beta - D_\beta z_\alpha
\nonumber
\end{eqnarray}
Now as the structure of Lagrangians and gauge transformations is the
same as in the flat case all variations with three derivatives
cancel each other leaving us with terms containing one derivative
only (and proportional to cosmological constant):
\begin{eqnarray}
\delta_0 {\cal L}_0 &=& - 3 \Omega \chi^{\mu\nu,\alpha} [ (3d-8)
(D T)_{\mu\nu,\alpha} - (2d-3) D_\alpha T_{\mu\nu} - 2(2d-3) g_{\nu\alpha}
(D T)_\mu ] - \nonumber \\
 && - 4 \Omega \eta^{\mu\nu\alpha} [ 4 D^\beta T_{\mu\nu\alpha,\beta}
 + 3(d-9) D_\mu T_{\nu\alpha} ] + 9 \Omega (d-3) z^{\alpha\beta}
 (D C)_{\alpha\beta} + \nonumber \\
 && + 2 \Omega x^{\alpha\beta} [ (2d-3) D^\mu \Phi_{\mu\alpha,\beta}
 + d D_\alpha \Phi_\beta - d g_{\alpha\beta} (D \Phi) ] + \nonumber \\
 && + 3 \Omega y^{\alpha\beta} [ 3 D^\mu \Phi_{\alpha\beta,\mu} - 2 (2d-6)
 D_\alpha \Phi_\beta ]
\end{eqnarray}
These terms do not contribute to calculations of variations with two
derivatives, so we keep the same structure of the Lagrangian terms
with one derivative:
\begin{eqnarray}
{\cal L}_1 &=& \alpha_1 [ T_{\mu\nu\alpha,\beta} D^\mu
\Phi^{\nu\alpha,\beta} - T_{\mu\nu} D_\alpha \Phi^{\mu\nu,\alpha}
- 2 T_{\mu\nu} D^\mu \Phi^\nu ] + \nonumber \\
 && + \frac{\alpha_2}{3} [ T_{\mu\nu\alpha,\beta} D^\beta
 C^{\mu\nu\alpha} - 3 T_{\mu\nu} (D C)^{\mu\nu} ] - \nonumber \\
 && - \frac{\alpha_5}{2} [ \Phi_{\mu\nu,\alpha} D^\alpha
 B^{\mu\nu} + 2 \Phi_\mu (D B)^\mu ] - \alpha_6 C_{\mu\nu\alpha}
 D^\mu B^{\nu\alpha}
\end{eqnarray}
as well as the same structure of non-derivative transformations for
all fields:
\begin{eqnarray}
\delta_1 T_{\mu\nu\alpha,\beta} &=& \alpha_3 [ g_{\alpha\beta} y_{\mu\nu}
- g_{\nu\beta} y_{\mu\alpha} + g_{\mu\beta} y_{\nu\alpha} ] +
\nonumber \\
 && + \alpha_4 [ g_{\alpha\beta} z_{\mu\nu}
- g_{\nu\beta} z_{\mu\alpha} + g_{\mu\beta} z_{\nu\alpha} ] \nonumber \\
\delta_1 \Phi_{\mu\nu,\alpha} &=& \alpha_1 \chi_{\mu\nu,\alpha} + \alpha_7
(g_{\nu\alpha} z_\mu - g_{\mu\alpha} z_\nu) \\
\delta_1 C_{\mu\nu\alpha} &=& \alpha_2 \eta_{\mu\nu\alpha} \qquad
\delta_1 B_{\alpha\beta} = \alpha_5 y_{\alpha\beta} + \alpha_6 z_{\alpha\beta}
\nonumber
\end{eqnarray}
In this, all variations with two derivatives indeed cancel each other
provided:
$$
\alpha_3 = \frac{2\alpha_1}{3(d-4)}, \qquad \alpha_4 = \frac{\alpha_2}{3(d-4)},
\qquad \alpha_7 = - \frac{\alpha_5}{2(d-3)}
$$
and we obtain non-derivative terms only:
\begin{eqnarray}
\delta_0 {\cal L}_1 + \delta_1 {\cal L}_0 &=& - \Omega \alpha_1
\chi^{\mu\nu,\alpha} [ (2d-3) \Phi_{\mu\nu,\alpha} - 2 d
g_{\nu\alpha} \Phi_\mu ] + \Omega \alpha_2 (d-3) \eta^{\mu\nu\alpha}
C_{\mu\nu\alpha} + \nonumber \\
 && + \Omega [ 2 \alpha_1 (d-5) - 6 \alpha_3 (d-1)] y^{\mu\nu}
 T_{\mu\nu} - 3 \Omega \alpha_5 (d-2) y^{\mu\nu} B_{\mu\nu} -
 \nonumber \\
 && - 2 \Omega [ \alpha_2 (d-2) + 3 \alpha_4 (d-1)] z^{\mu\nu}
 T_{\mu\nu} + \Omega [2 \alpha_7 d - \alpha_5 (d-1)] z^\mu \Phi_\mu
\end{eqnarray}

At last we add to the Lagrangian the most general mass-like
terms for all fields:
\begin{eqnarray}
{\cal L}_2 &=& \frac{c_1}{2} T^{\mu\nu\alpha,\beta} T_{\mu\nu\alpha,\beta}
+ \frac{c_2}{2} T^{\mu\nu} T_{\mu\nu} + c_3 T^{\mu\nu} B_{\mu\nu}
+ \frac{c_4}{2} B^{\mu\nu} B_{\mu\nu} + \nonumber \\
 && + \frac{c_5}{2} \Phi^{\mu\nu,\alpha} \Phi_{\mu\nu,\alpha}
 + \frac{c_6}{2} \Phi^\mu \Phi_\mu + \frac{c_7}{2} C^{\mu\nu\alpha}
 C_{\mu\nu\alpha}
\end{eqnarray}
and require the cancellation of all variations with one derivative
(including $\delta_0 {\cal L}_0$) and without derivatives (taking
into account $\delta_0 {\cal L}_1 + \delta_1 {\cal L}_0$). This allows
us to express all the parameters in the Lagrangian and gauge
transformations in terms of $\alpha_1$ and $\alpha_2$
$$
\alpha_5{}^2 = \frac{4(d-3)}{3(d-4)} \alpha_1{}^2 + 12 \Omega (d-3), \qquad
\alpha_6{}^2 = \frac{d-3}{3(d-4)} \alpha_2{}^2 - 12 \Omega (d-3)
$$
$$
c_1 = - \frac{\alpha_1{}^2}{3} - \Omega (3d-8), \qquad c_2 = \alpha_1{}^2
+ 3 \Omega (2d-3), \qquad c_3 = - \frac{\alpha_1 \alpha_2}{2}
$$
$$
c_4 = \frac{d-2}{4(d-3)} \alpha_5{}^2, \qquad c_5 = \Omega (2d-3), \qquad
c_6 = - 2 \Omega d, \qquad c_7 = - \Omega (d-3)
$$
and gives us an important relation on this parameters:
\begin{equation}
4 \alpha_1{}^2 - \alpha_2{}^2 + 36 \Omega (d-4) = 0
\end{equation}

Thus we have a one parameter family of Lagrangians. But it is not
evident which parameter or combination of parameters should be
identified with mass because the last relation means that
exactly as in the previous case for the nonzero value of
cosmological constant it is not possible to set $\alpha_1 = 0$
and $\alpha_2 = 0$ simultaneously (recall that $d \ne 4$).
So there is no fully massless limit in (A)dS but there are two
partially massless ones.

In Anti de Sitter space ($\Omega < 0$) one can set $\alpha_2 = 0$.
As a result the whole system decompose into two subsystems. One of
them with the fields $T_{\mu\nu\alpha,\beta}$ and $\Phi_{\mu\nu,\alpha}$
with the Lagrangian
\begin{eqnarray}
{\cal L} &=& {\cal L}_0 (T_{\mu\nu\alpha,\beta}) + {\cal L}_0
(\Phi_{\mu\nu,\alpha}) + \alpha_1 [ T_{\mu\nu\alpha,\beta} D^\mu
\Phi^{\nu\alpha,\beta} - T_{\mu\nu} D_\alpha \Phi^{\mu\nu,\alpha}
- 2 T_{\mu\nu} D^\mu \Phi^\nu ] - \nonumber \\
 && - 2 \Omega T^{\mu\nu\alpha,\beta} T_{\mu\nu\alpha,\beta} -
 \frac{3}{2} \Omega (d-9) T^{\mu\nu} T_{\mu\nu} + \frac{1}{2}
 \Omega (2d-3) \Phi^{\mu\nu,\alpha} \Phi_{\mu\nu,\alpha} -
 \Omega d \Phi^\mu \Phi_\mu
\end{eqnarray}
where $\alpha_1 = 3\sqrt{-\Omega(d-4)}$, which is invariant under
the following gauge transformations
\begin{eqnarray}
\delta T_{\mu\nu\alpha,\beta} &=& 3 D_\beta \eta_{\mu\nu\alpha}
+ D_\alpha \eta_{\mu\nu\beta} + D_\mu \eta_{\nu\alpha\beta}
- D_\nu \eta_{\mu\alpha\beta} + \nonumber \\
 && + D_\mu \chi_{\nu\alpha,\beta} - D_\nu
\chi_{\mu\alpha,\beta} + D_\alpha \chi_{\mu\nu,\beta} + \nonumber \\
 && + \frac{2 \alpha_1}{3(d-4)} [ g_{\alpha\beta} y_{\mu\nu}
 - g_{\nu\beta} y_{\mu\alpha} + g_{\mu\beta} y_{\nu\alpha} ] + \nonumber \\
\delta \Phi_{\mu\nu,\alpha} &=& D_\mu x_{\nu\alpha} - D_\nu
x_{\mu\alpha} + 2 D_\alpha y_{\mu\nu} - D_\mu y_{\nu\alpha}
+ D_\nu y_{\mu\alpha} + \alpha_1 \chi_{\mu\nu,\alpha}
\end{eqnarray}
gives us one more example of partially massless theory.

The other one with the fields $C_{\mu\nu\alpha}$ and $B_{\mu\nu}$ is
just gauge invariant description of massive third rank antisymmetric
tensor with the Lagrangian
\begin{equation}
{\cal L} = {\cal L}_0 (C_{\mu\nu\alpha}) + {\cal L}_0 (B_{\mu\nu}) - 2
\sqrt{- 3 \Omega (d-3)} C_{\mu\nu\alpha} D^\mu B^{\nu\alpha} -
\frac{\Omega (d-3)}{2} C^{\mu\nu\alpha} C_{\mu\nu\alpha}
\end{equation}
and gauge transformations
\begin{eqnarray}
\delta C_{\mu\nu\alpha} &=& D_\mu z_{\nu\alpha} - D_\nu z_{\mu\alpha}
+ D_\alpha z_{\mu\nu} \nonumber \\
\delta B_{\mu\nu} &=& D_\mu z_\nu - D_\nu z_\mu + 2 \sqrt{ 3 \Omega (d-3)}
z_{\mu\nu}
\end{eqnarray}

In the de Sitter space ($\Omega > 0$) one can set $\alpha_1 = 0$
instead. Once again the whole system breaks into two decoupled
subsystems. One of them give another partially massless theory
in terms of $T_{\mu\nu\alpha,\beta}$ and $C_{\mu\nu\alpha}$ with
the Lagrangian
\begin{eqnarray}
{\cal L} &=& {\cal L}_0 (T_{\mu\nu\alpha,\beta}) + {\cal L}_0
(C_{\mu\nu\alpha}) + \frac{\alpha_2}{3} [ T_{\mu\nu\alpha,\beta}
D^\beta C^{\mu\nu\alpha} - 3 T_{\mu\nu} (D C)^{\mu\nu} ] - \nonumber \\
 && - 2 \Omega T^{\mu\nu\alpha,\beta} T_{\mu\nu\alpha,\beta} -
 \frac{3}{2} \Omega (d-9) T^{\mu\nu} T_{\mu\nu} -
 \frac{\Omega (d-3)}{2} C^{\mu\nu\alpha} C_{\mu\nu\alpha}
\end{eqnarray}
where $\alpha_2 = 6\sqrt{\Omega(d-4)}$, which is invariant under
\begin{eqnarray}
\delta T_{\mu\nu\alpha,\beta} &=& 3 D_\beta \eta_{\mu\nu\alpha}
+ D_\alpha \eta_{\mu\nu\beta} + D_\mu \eta_{\nu\alpha\beta}
- D_\nu \eta_{\mu\alpha\beta} + \nonumber \\
 && + D_\mu \chi_{\nu\alpha,\beta} - D_\nu
\chi_{\mu\alpha,\beta} + D_\alpha \chi_{\mu\nu,\beta} + \nonumber \\
 && + \frac{\alpha_2}{3(d-4)} [ g_{\alpha\beta} z_{\mu\nu}
 - g_{\nu\beta} z_{\mu\alpha} + g_{\mu\beta} z_{\nu\alpha} ] + \\
\delta C_{\mu\nu\alpha} &=& D_\mu z_{\nu\alpha} - D_\nu
z_{\mu\alpha} + D_\alpha z_{\mu\nu} + \alpha_2 \eta_{\mu\nu\alpha} \nonumber
\end{eqnarray}

The rest fields $\Phi_{\mu\nu,\alpha}$ and $B_{\mu\nu}$ give exactly
the same partially massless theory that we have in the previous section
for the de Sitter space.

So the general pattern of such massive theory both in Minkowski space
as well as in (A)dS resembles very much the one for the field
$\Phi_{\mu\nu,\alpha}$ considered in the previous section. It is not
hard to construct a generalization of this theory to the case of
the tensor like $T_{[\mu_1\mu_2\dots\mu_n],\nu}$ for arbitrary $n$.
But more general case like $R_{[\mu_1\dots\mu_n],[\nu_1\dots\nu_m]}$
contains special symmetric case $m=n$ which requires separate study.
The most simple (but may be the most interesting) example --- tensor
$R_{[\mu\nu],[\alpha\beta]}$ which will be the subject of our next
section.

\section{$R_{[\mu\nu],[\alpha\beta]}$ tensor}

Our last example --- tensor $R_{[\mu\nu],[\alpha\beta]}$ antisymmetric
on both first as well as second pair of indices and satisfying
$R_{\mu\nu,\alpha\beta} = R_{\alpha\beta,\mu\nu}$ and
$R_{[\mu\nu,\alpha]\beta} = 0$. We start with the massless case in the
Minkowski space and consider the Lagrangian
\begin{eqnarray}
{\cal L}_0 &=& \frac{1}{8} \partial^\rho R^{\mu\nu,\alpha\beta}
\partial_\rho R_{\mu\nu,\alpha\beta} - \frac{1}{2}
(\partial R)^{\nu,\alpha\beta} (\partial R)_{\nu,\alpha\beta} -
(\partial R)^{\nu,\alpha\beta} \partial_\beta R_{\nu\alpha} -
\nonumber \\
 && - \frac{1}{2} \partial^\rho R^{\mu\nu} \partial_\rho R_{\mu\nu} +
 (\partial R)^\mu (\partial R)_\mu - \frac{1}{2} (\partial R)^\mu
 \partial_\mu R + \frac{1}{8} \partial^\mu R \partial_\mu R
\end{eqnarray}
It is easy to check that this Lagrangian is invariant under the
following gauge transformations:
\begin{equation}
\delta R_{\mu\nu,\alpha\beta} = \partial_\mu \chi_{\nu,\alpha\beta}
- \partial_\nu \chi_{\mu,\alpha\beta} + \partial_\alpha
\chi_{\beta,\mu\nu} - \partial_\beta \chi_{\alpha,\mu\nu}
\end{equation}
where parameter $\chi_{\mu,[\alpha\beta]}$ antisymmetric on the two
indices and satisfies $\chi_{[\mu,\alpha\beta]} = 0$. Note that due to
symmetry property $R_{\mu\nu,\alpha\beta} = R_{\alpha\beta,\mu\nu}$
we have only one gauge transformation instead of two as for the
previous cases. This gauge transformation is also reducible because
if one set
\begin{equation}
\chi_{\mu,\alpha\beta} = 2 \partial_\mu y_{\alpha\beta} - \partial_\alpha
y_{\beta\mu} + \partial_\beta y_{\alpha\mu}
\end{equation}
where $y_{\alpha\beta}$ antisymmetric tensor, then $R_{\mu\nu,\alpha\beta}$
remains invariant.

Following our general procedure we introduce one Goldstone field
$\Phi_{[\mu\nu],\alpha}$ with the same symmetry properties as
$\chi_{\mu,\alpha\beta}$ with the same massless Lagrangian and its
own gauge transformations as before. Then by adding the following
low derivatives terms to the sum of massless Lagrangians
\begin{eqnarray}
{\cal L}_1 &=& m [ R^{\mu\nu,\alpha\beta} \partial_\mu \Phi_{\nu,\alpha\beta}
- 2 R^{\mu\nu} (\partial \Phi)_{\mu,\nu} - 2 R^{\mu\nu} \partial_\mu
\Phi_\nu + R (\partial \Phi) ] - \nonumber \\
 && - \frac{m^2}{8} [ R^{\mu\nu,\alpha\beta} R_{\mu\nu,\alpha\beta}
 - 4 R^{\mu\nu} R_{\mu\nu} + R^2]
\end{eqnarray}
we can still have gauge invariance under the $\chi_{\mu,\alpha\beta}$
transformations for massive field provided
\begin{equation}
\delta_1 \Phi_{\alpha\beta,\mu} = m \chi_{\mu,\alpha\beta}
\end{equation}

Recall that our Goldstone field $\Phi_{\mu\nu,\alpha}$ has its own
gauge transformations with the parameters $x_{\{\alpha\beta\}}$ and
$y_{[\alpha\beta]}$. Due to reducibility of gauge transformations
for $R_{\mu\nu,\alpha\beta}$ it turns out that one has to introduce
one new Goldstone field $h_{\{\alpha\beta\}}$ only. Indeed with the
additional terms
\begin{eqnarray}
\Delta {\cal L} &=& {\cal L}_0 (h_{\alpha\beta}) - m \alpha_2 [
\Phi_{\mu\nu,\alpha} \partial^\mu h^{\nu\alpha} + \Phi_\mu
(\partial h)^\mu - \Phi_\mu \partial^\mu h ] + \nonumber \\
 && + \frac{m^2}{2} [ - \alpha_2 R^{\mu\nu} H_{\mu\nu} +
\frac{\alpha_2}{2} R h + \frac{d-2}{d-4} (h^{\mu\nu} h_{\mu\nu}
- h^2) ]
\end{eqnarray}
our theory becomes invariant not only under $\chi_{\mu,\alpha\beta}$
transformations, but under $x_{\alpha\beta}$ and $y_{\alpha\beta}$
ones as well, provided
\begin{eqnarray}
\delta_1 R_{\mu\nu,\alpha\beta} &=& \frac{2m}{d-4} ( g_{\mu\alpha}
x_{\nu\beta} - g_{\mu\beta} x_{\nu\alpha} - g_{\nu\alpha} x_{\mu\beta}
+ g_{\nu\beta} x_{\mu\alpha} ) \nonumber \\
\delta_1 h_{\alpha\beta} &=& m \alpha_2 x_{\alpha\beta}
\end{eqnarray}
Here $\alpha_2 = 2 \sqrt{\frac{d-3}{d-4}}$. Once again, the whole
construction works for $d \ge 5$ only.

Moreover, due to the reducibility of gauge transformations of the
field $\Phi_{\mu\nu,\alpha}$ the Lagrangian obtained turns out to be
invariant under one more gauge transformation with vector parameter
\begin{eqnarray}
\delta \Phi_{\mu\nu,\alpha} &=& - \frac{m}{\sqrt{(d-3)(d-4)}} (g_{\nu\alpha}
x_\mu - g_{\mu\alpha} x_\nu) \nonumber \\
\delta h_{\alpha\beta} &=& \partial_\alpha x_\beta + \partial_\beta
x_\alpha
\end{eqnarray}

So the whole massive theory requires three fields $R_{\mu\nu,\alpha\beta}$,
$\Phi_{\mu\nu,\alpha}$ and $h_{\alpha\beta}$ only and has four gauge
transformations with the parameters $\chi_{\mu,\alpha\beta}$,
$x_{\alpha\beta}$, $y_{\alpha\beta}$ and $x_\alpha$.

Let us turn to the (A)dS case. As we have already noted our main field
$R_{\mu\nu,\alpha\beta}$ has one gauge transformation only, so one may
expect that there exist fully massless theory for this field in (A)dS. 
Indeed if we consider the Lagrangian ${\cal L}_0 (R_{\mu\nu,\alpha\beta})$ 
and gauge transformations with the parameter $\chi_{\mu,\alpha\beta}$ where
all the derivatives are replaced by the covariant ones, then the
variation of such Lagrangian:
\begin{equation}
\delta_0 {\cal L}_0 = - \Omega [ (d+2) \chi^{\mu,\alpha\beta}
(D R)_{\nu,\alpha\beta} + 10 \chi^{\alpha,\beta\nu} D_\nu R_{\alpha\beta}
+ 10 \chi^\alpha (D R)_\alpha + (d-8) \chi^\alpha D_\alpha R ]
\end{equation}
could be perfectly canceled by the addition of the following mass-like
terms:
\begin{equation}
{\cal L}_m = - \frac{\Omega}{8} [ (d+2) R^{\mu\nu,\alpha\beta}
R_{\mu\nu,\alpha\beta} - 20 R^{\mu\nu} R_{\mu\nu} + (d-8) R^2 ]
\end{equation}

Now let us consider massive case. This time (due to existence of
fully massless limit) we will follow the same convention that we
use in the case of completely symmetric tensor fields \cite{Zin01}, namely
we will call "mass" the parameter which would be the mass in the flat
space limit. So we introduce two additional fields $\Phi_{\mu\nu,\alpha}$
and $h_{\alpha\beta}$ with their own massless Lagrangians and gauge
transformations and add to the sum of massless Lagrangians the following
terms with one derivative:
\begin{eqnarray}
{\cal L}_1 &=& m [ R^{\mu\nu,\alpha\beta} D_\mu \Phi_{\nu,\alpha\beta}
- 2 R^{\mu\nu} (D \Phi)_{\mu,\nu} - 2 R^{\mu\nu} D_\mu \Phi_\nu
+ R (D \Phi) ] - \nonumber \\
 && - \alpha_2 [ \Phi_{\mu\nu,\alpha} D^\mu h^{\nu\alpha} + \Phi_\mu
 (D h)^\mu - \Phi_\mu D^\mu h ]
\end{eqnarray}
as well as the following non-derivative terms to the transformation laws:
\begin{eqnarray}
\delta_1 R_{\mu\nu,\alpha\beta} &=& \frac{2m}{d-4} ( g_{\mu\alpha}
x_{\nu\beta} - g_{\mu\beta} x_{\nu\alpha} - g_{\nu\alpha} x_{\mu\beta}
+ g_{\nu\beta} x_{\mu\alpha} ) \nonumber \\
\delta_1 \Phi_{\alpha\beta,\mu} &=& m \chi_{\mu,\alpha\beta}
- \frac{\alpha_2}{2(d-3)} (g_{\beta\mu} x_\alpha - g_{\alpha\mu}
x_\beta) \\
\delta_1 h_{\alpha\beta} &=& \alpha_2 x_{\alpha\beta} \nonumber
\end{eqnarray}
Then straightforward calculations show that all variations with two
derivatives cancel each other leaving us with:
\begin{eqnarray}
\delta_0 {\cal L}_1 + \delta_1 {\cal L}_0 &=& - \Omega [ 3m
\chi^{\mu,\alpha\beta} \Phi_{\alpha\beta,\mu} - 2m (d-6)
\chi^\mu \Phi_\mu + \nonumber \\
 && + \frac{2m(d-3)}{d-4} (d x^{\alpha\beta} R_{\alpha\beta}
 - x R ) - \alpha_2 (d x^{\alpha\beta} h_{\alpha\beta} - x h) ]
\end{eqnarray}

Now if we add the most general mass-like terms for all three fields:
\begin{eqnarray}
{\cal L}_2 &=& \frac{c_1}{8} R^{\mu\nu,\alpha\beta} R_{\mu\nu,\alpha\beta}
+ \frac{c_2}{4} R^{\mu\nu} R_{\mu\nu} + \frac{c_3}{8} R^2 +
\frac{c_4}{2} \Phi^{\mu\nu,\alpha} \Phi_{\mu\nu,\alpha}
+ \frac{c_5}{2} \Phi^\mu \Phi_\mu + \nonumber \\
 && + c_6 R^{\mu\nu} h_{\mu\nu} + c_7 R h + \frac{c_8}{2}
 h^{\mu\nu} h_{\mu\nu} + \frac{c_9}{2} h^2
\end{eqnarray}
and require cancellation of all variations then we obtain the following
expressions for the parameters
$$
c_1 = - m^2 - \Omega (d+2), \qquad c_2 = 2 m^2 + 10 \Omega, \qquad
c_3 = - m^2 + \Omega (d-8),
$$
$$
c_4 = 6 \Omega, \qquad c_5 = 2 \Omega (d-6), \qquad c_6 = -
\frac{m \alpha_2}{2}, \qquad c_7 = \frac{m\alpha_2}{4},
$$
$$
c_8 = \frac{d-2}{d-3} \alpha_2{}^2 - 2 \Omega, \qquad
c_9 = - \frac{d-2}{d-3} \alpha_2{}^2 - \Omega (d-3)
$$
as well as the following relation on the main parameter $\alpha_2$
\begin{equation}
\alpha_2{}^2 = \frac{4(d-3)}{d-4} [ m^2 - \Omega (d-4)]
\end{equation}

In (A)dS one can set $m = 0$. In this the field $R_{\mu\nu,\alpha\beta}$
decouples and gives fully massless theory, while the others ---
$\Phi_{\mu\nu,\alpha}$ and $h_{\alpha\beta}$ gives the same partially
massless theory as in the first section. At the same time in the de Sitter
space we have unitary forbidden region, because the last relation
requires $m^2 \ge \Omega (d-4)$. The boundary of this region gives
($\alpha_2 = 0$) partially massless theory with the fields
$R_{\mu\nu,\alpha\beta}$ and $\Phi_{\mu\nu,\alpha}$ with the Lagrangian
\begin{eqnarray}
{\cal L} &=& {\cal L}_0 (R_{\mu\nu,\alpha\beta}) + {\cal L}_0
(\Phi_{\mu\nu,\alpha}) + \nonumber \\
 && + m [ R^{\mu\nu,\alpha\beta} D_\mu \Phi_{\nu,\alpha\beta}
 - 2 R^{\mu\nu} (D \Phi)_{\mu,\nu} - 2 R^{\mu\nu} D_\mu \Phi_\nu
 + R (D \Phi) ] - \nonumber \\
 && - \frac{\Omega}{4} [ (d-1) R^{\mu\nu,\alpha\beta}
 R_{\mu\nu,\alpha\beta} - 2 (d+1) R^{\mu\nu} R_{\mu\nu}
 + 2 R^2 - \nonumber \\
 && - 6 \Phi^{\mu\nu,\alpha} \Phi_{\mu\nu,\alpha} - 4 (d-6) \Phi^\mu
 \Phi_\mu ]
\end{eqnarray}
which is invariant under the following gauge transformations:
\begin{eqnarray}
\delta R_{\mu\nu,\alpha\beta} &=& D_\mu \chi_{\nu,\alpha\beta}
- D_\nu \chi_{\mu,\alpha\beta} + D_\alpha \chi_{\beta,\mu\nu}
- D_\beta \chi_{\alpha,\mu\nu} + \nonumber \\
 && + \frac{2m}{d-4} ( g_{\mu\alpha} x_{\nu\beta} - g_{\mu\beta}
 x_{\nu\alpha} - g_{\nu\alpha} x_{\mu\beta} + g_{\nu\beta}
 x_{\mu\alpha} ) \\
\delta \Phi_{\mu\nu,\alpha} &=& D_\mu x_{\nu\alpha} - D_\nu x_{\mu\alpha}
+ 2 D_\alpha y_{\mu\nu} - D_\mu y_{\nu\alpha} + D_\nu y_{\mu\alpha}
+ m \chi_{\alpha,\mu\nu} \nonumber
\end{eqnarray}
In this, $h_{\alpha\beta}$ just describes the usual massless spin-2
particle.

\section*{Conclusion}

In this paper we have managed to construct gauge invariant formulations
for massive particles described by mixed symmetry tensors
$\Phi_{[\mu\nu],\alpha}$, $T_{[\mu\nu\alpha],\beta}$ and
$R_{[\mu\nu],[\alpha\beta]}$. In all three cases it was crucial
to take into account the reducibility of corresponding gauge
transformations to determine appropriate set of Goldstone fields.
We have seen that such formulations admit smooth deformation to
the (Anti) de Sitter space without introduction of any additional
fields. This, in turn, allows us to investigate possible massless
as well as partially massless limits for such theories. Our results
agree with the observations in \cite{BMV00} and give a number of
new examples of partially massless theories both in de Sitter as
well as in Anti de Sitter spaces. Here we did not try to consider
generalizations to the more general mixed symmetry tensors, but
we hope that three explicit examples constructed provide a good
starting point for such generalizations.

\end{document}